\documentclass[twocolumn]{aastex63}

\usepackage{natbib}
\usepackage{hyperref}

\usepackage{lineno}

\def\deg{\ifmmode^\circ\else$^\circ$\fi}
\def\kps{km\thinspace s$^{-1}$}

\newcommand{\Msolar}{M$_{\odot}$}

\newcommand{\mic}{\,$\mu$m}

\newcommand{\micro}{$\mu$m}
\newcommand{\simi}{$\sim$}
\newcommand{\as}{$''$}

\newcommand{\htwo}{H\,{\sc ii}}
\newcommand{\cone}{Class\,{\sc I} YSOs}

\newcommand{\tco}{$^{13}$CO(J = 2--1)}
\newcommand{\twco}{$^{12}$CO(J = 1--0)}

\shorttitle{CCC and HFSs in W31}
\shortauthors{A.~K.  Maity et al.}

\begin{document}

\title{Unraveling the observational signatures of cloud-cloud collision and hub-filament systems in W31}

\correspondingauthor{A.~K.  Maity}
\email{Email: arupmaity@prl.res.in}

\author[0000-0002-7367-9355]{A.~K. Maity}
\affiliation{Astronomy \& Astrophysics Division, Physical Research Laboratory, Navrangpura, Ahmedabad 380009, India}
\affiliation{Indian Institute of Technology Gandhinagar Palaj, Gandhinagar 382355, India}

\author[0000-0001-6725-0483]{L.~K. Dewangan}
\affiliation{Astronomy \& Astrophysics Division, Physical Research Laboratory, Navrangpura, Ahmedabad 380009, India}

\author[0000-0003-2062-5692]{H.~Sano}
\affiliation{National Astronomical Observatory of Japan, Mitaka, Tokyo 181-8588, Japan}

\author[0000-0002-1411-5410]{K.~Tachihara}
\affiliation{Department of Physics, Nagoya University, Furo-cho, Chikusa-ku, Nagoya 464-8601, Japan}

\author{Y.~Fukui}
\affiliation{Department of Physics, Nagoya University, Furo-cho, Chikusa-ku, Nagoya 464-8601, Japan}

\author[0000-0001-8812-8460]{N.~K.~Bhadari}
\affiliation{Astronomy \& Astrophysics Division, Physical Research Laboratory, Navrangpura, Ahmedabad 380009, India}
\affiliation{Indian Institute of Technology Gandhinagar Palaj, Gandhinagar 382355, India}

%
%


\begin{abstract}
To understand the formation process of massive stars, we present a multi-scale and multi-wavelength study of the W31 complex hosting two extended {\htwo} regions (i.e., G10.30-0.15 (hereafter, W31-N) and G10.15-0.34 (hereafter, W31-S)) powered by a cluster of O-type stars.  Several Class\,{\sc I} protostars and a total of 49 ATLASGAL 870 $\mu$m dust clumps (at d = 3.55 kpc) are found toward the {\htwo} regions where some of the clumps are associated with the molecular outflow activity. These results confirm the existence of a single physical system hosting the early phases of star formation.  The {\it Herschel} 250 {\micro} continuum map shows the presence of hub-filament system (HFS) toward both W31-N and W31-S. The central hubs harbour {\htwo} regions and they are depicted with extended structures (with T$_{\text{d}}$ $\sim$ 25--32~K) in the {\it Herschel} temperature map. In the direction of W31-S, an analysis of the NANTEN2 {\twco} and SEDIGISM {\tco} line data supports the presence of two cloud components around 8 and 16 {\kps}, and their connection in velocity space.  A spatial complementary distribution between the two cloud components is also investigated toward W31-S, where the signposts of star formation, including massive O-type stars, are concentrated. These findings favor the applicability of cloud-cloud collision (CCC) around {\simi}2 Myr ago in W31-S. Overall, our observational findings support the theoretical scenario of CCC in W31, which explains the formation of massive stars and the existence of HFSs.
\end{abstract}
%
%
\section{Introduction}
\label{sec:intro}
Massive OB stars ($\gtrsim$ 8 M$_{\odot}$) have a great impact on the galaxy structure, evolution, and next-generation star formation. However, the formation of such stars is not fully understood \citep[e.g.,][]{zinnecker07,tan14,motte_2018}. The study of the formation of massive OB stars is related to the understanding of the mass accumulation processes from the surroundings \citep[e.g.,][]{motte_2018,hirota18}. In this context, the core-fed scenario and the clump-fed scenario have been proposed \citep[e.g.,][]{rosen20}. In the core-fed scenario, massive stars are formed by the collapse of massive prestellar cores \citep[i.e., monolithic collapse model;][]{mckee_2003}. In the clump-fed scenario, massive stars are born through inflow material from very large scales of 1--10 pc, which can be funneled via the molecular cloud filaments \citep{bonnell_2001,bonnell_2004,bonnell_2006,vazquez_2009,vazquez_2017,vazquez_2019,padoan_2020}. 
Several recent observational research works have shown that the sites of protocluster hosting early phases of massive stars are often found at the junctions of dust and molecular filaments (i.e., hub-filament systems (HFSs); \citet[]{myers_2009}), which play a pivotal role in the process of mass accumulation \citep[e.g.,][]{tige_2017,motte_2018,trevino19}. 
Additionally, cloud-cloud collision \citep[CCC; see][for a comprehensive review]{fukui21} process has received great attention 
over the core-fed and clump-fed scenarios. The CCC produces a shock-compressed layer at common zones of colliding clouds, where one can expect an enhancement of gas density and effective sound speed 
\citep{habe92,anathpindika_2010,inoue_2013,haworth_2015,haworth_2015b,torri_2017,takahira_2018}. 
Such an environment is believed to be appropriate for the formation of massive OB stars.  Furthermore, the shock-compressed layer produced by the colliding clouds can fragment into filaments, which form the observed pattern of filaments \citep[e.g., hub or spokes systems;][]{balfour15,inoue18}. However, there are very limited observational works available in the literature to investigate the connection between CCC and HFSs in potential massive star-forming sites 
\citep[e.g.,][]{fukui19ex,tokuda19ex,beltran22,dewangan22ex}. In this relation, the present paper deals with an extended star-forming complex W31, hosting several massive O-type stars.

The W31 complex is known to host two well-studied {\htwo} regions \citep{Wood,kim01}, which are G10.3-0.15 (hereafter, W31-N) and G10.15-0.34 (hereafter, W31-S). Both the {\htwo} regions were well examined using radio continuum maps, and molecular line data sets \citep{W31Kim2002,Beuther}.  The W31-N {\htwo} region associated with a mid-infrared (MIR) bubble CN 148 is excited by O-type stars \citep{bik_2005, CN148} and the location of one of these exciting O-type stars are highlighted in the top left panel of Figure 1 in \citet{Beuther}. 
Using a multi-wavelength approach, the physical environment of W31-N and star formation activities are extensively studied in \citet{CN148}.  The W31-S {\htwo} region emits more than 10$^{50}$ Lyman continuum photons per second, which favors the presence of several O-type stars \citep{Blum}.  Also, it is one of the largest {\htwo} regions in our galaxy with intense star formation activities \citep{moises_2011}.  Different distances (e.g., 3.4 kpc \citep{Blum}, 6 kpc \citep{wilson_1974,downes_1980} and 14.5 kpc \citep{corbel_1997}) to W31 have been reported in the literature. However, \citet{moises_2011} estimated an average K-band spectrophotometric distance of \simi 2.39 kpc and \simi 3.55 kpc for W31-N and W31-S, respectively.  In this paper, we choose a distance of 3.55 kpc for both the sites W31-N and W31-S, which is based on the measured distances of APEX Telescope Large Area Survey of the Galaxy \citep[ATLASGAL;][]{schuller_2009} 870 $\mu$m 
dust continuum clumps \citep{Atlasgal} associated with W31 (see Section \ref{dis_alt}).

Despite the availability of numerous studies, to our knowledge, no mechanisms are reported to successfully explain the birth of O-type stars in W31.  Hence, this paper aims to unveil the physical processes responsible for the formation of such massive stars in W31. 
In this context, a careful and detailed study of the dust continuum emission and the distribution of molecular gas is essential.  
Hence, multi-wavelength data have been carefully examined in W31 (see Table~\ref{tab1}).  In particular, to study the molecular gas, we analyzed new $^{12}$CO(1--0) line data from the NANTEN2 telescope and the publicly available $^{13}$CO(2--1) line data from the Structure, Excitation, and Dynamics of the Inner Galactic Interstellar Medium \citep[SEDIGISM;][]{schuller17} survey.

Section~\ref{sec:obser} presents the observational data sets discussed in this paper.  The outcomes of this paper are given in Section~\ref{sec:data}.  In Section~\ref{sec:disc}, the results of this study are thoroughly discussed to explain the birth process of massive stars.  Finally, Section~\ref{sec:conc} presents the summary and conclusions of this study.
\section{Data and Analysis}
\label{sec:obser}
\subsection {New molecular line data: NANTEN2 $^{12}$CO(J = 1--0)}
\label{nanten}
Observations of $^{12}$CO(J = 1--0) line emission at 115.27 GHz were conducted 
in 2012 December and 2013 January using the NANTEN2 millimeter (mm)/sub-mm radio telescope of Nagoya University installed at an altitude of 4865-m in the Atacama Desert in Chile. 
We mapped an area of 0$\degr$.7 $\times$ 0$\degr$.7 centered at W31 using on-the-fly 
mapping mode with Nyquist sampling. The front end was a double-sideband (DSB) heterodyne receiver 
equipped with a 4-K cooled Nb superconductor-insulator-superconductor mixer.
The back end was a digital Fourier-transform spectrometer with a bandwidth of 1 GHz 
and a channel spacing of 61 kHz, corresponding to a velocity coverage of $\sim$2600 {\kps} 
and a velocity resolution of $\sim$0.16 {\kps} at 115 GHz.  The typical system temperature 
was $\sim$190--240 K in the DSB, including the atmosphere. 
After convolution with a two-dimensional Gaussian function of 90$''$ (FWHM), we obtained a data cube 
with a beam size of $\sim$180$''$ (FWHM). 
The absolute intensity was calibrated by observing Orion-KL [$\alpha_{2000}$ = 05$^{h}$35$^{m}$14\rlap.$^{s}$48; $\delta_{2000}$ = $-$05$\degr$22$'$27\rlap.{$''$}55] 
and IRAS 16239-2422 [$\alpha_{2000}$ = 16$^{h}$32$^{m}$23\rlap.$^{s}$3; $\delta_{2000}$ = $-$24$\degr$28$'$39\rlap.{$''$}2] \citep{ridge06}.
The pointing accuracy was checked every two hours to achieve an offset within 2$''$, 
verified by observing IRC+10216 and the edge of the Sun. 
The typical noise fluctuation is $\sim$1.0 K at the velocity resolution of 0.16 {\kps}.
%
\subsection {Archive datasets}
\label{datasets}
In this paper, we used several data sets obtained from different publicly available surveys, which are listed in Table~\ref{tab1}. In the direction of our selected target area, dust continuum emission was examined using the ATLASGAL 870 $\mu$m continuum map and the {\it Herschel} maps at 250--500 $\mu$m. We obtained the {\it Herschel} temperature map (resolution $\sim$12$''$) and the H$_2$ column density map (resolution $\sim$12$''$) from a publicly available data  base\footnote[1]{http://www.astro.cardiff.ac.uk/research/ViaLactea/}, which were produced by the application of PPMAP algorithm \citep{marsh_2015,marsh_2017} to the {\it Herschel} maps at 70--500 $\mu$m from the Hi-GAL survey \citep{Molinari10b}. 
From \citet{Atlasgal}, the physical parameters (such as velocity and distance) of the ATLASGAL dust continuum clumps at 870 $\mu$m  were collected in our selected target complex.  The information of the ATLASGAL clumps associated with the $^{13}$CO outflows was taken from \citet{Yang2022}.  The NRAO VLA Sky Survey (NVSS) 1.4 GHz continuum map was examined to explore the ionized emission. 

We have utilized the $^{13}$CO(J = 2--1) line data \citep[beam size $\sim$30$''$; rms $\sim$0.8--1.0~K;][]{schuller17} from the SEDIGISM survey \citep[see][for more details]{Schuller2021}.
We have smoothed the obtained SEDIGISM $^{13}$CO(J = 2--1) line data with a Gaussian function having a width of 3 pixels.  With a pixel-scale of 9\rlap.{$''$}5, the resultant angular resolution becomes $\sqrt{30^2 + 28.5^2} \approx$ 41\rlap.{$''$}4 after the Gaussian smoothing process.  We obtained the photometric magnitude of point-likes sources at 3.6, 4.5, and 5.8 $\mu$m from the {\it Spitzer}-GLIMPSE survey.  The GLIMPSE-I Spring $' $07 highly reliable photometric catalog was used to identify protostars in this paper. 
\begin{table*}
\scriptsize
\setlength{\tabcolsep}{0.005in}
\centering
\caption{List of archive data sets utilized in this work.}
\label{tab1}
\begin{tabular}{lcccr}
\hline 
  Survey  &  Wavelength/Frequency/line(s)       &  Resolution ($\arcsec$)        &  Reference \\   
\hline
NRAO VLA Sky Survey (NVSS)       & 1.4 GHz  & $\sim$45 & \citet{NVSS}\\
SEDIGISM &  $^{13}$CO(J = 2--1) & $\sim$30        &\citet{schuller17}\\
ATLASGAL                 &870 $\mu$m                     & $\sim$19        & \citet{schuller_2009} \\
{\it Herschel} Infrared Galactic Plane Survey (Hi-GAL)                              & 250, 350, 500 $\mu$m     & $\sim$18, 25, 37         &\citet{Molinari10a}\\
{\it Spitzer} Galactic Legacy Infrared Mid-Plane Survey Extraordinaire (GLIMPSE) & 3.6, 4.5, 5.8 $\mu$m & $\sim$2 & \citet{Benjamin}\\
\hline          
\end{tabular}			
\end{table*}			
%
\section{Results}
\label{sec:data}
\subsection{Dust clumps, ionized regions, and embedded protostars in W31}
\label{dis_alt}
\begin{figure*}
\epsscale{1}
\plotone{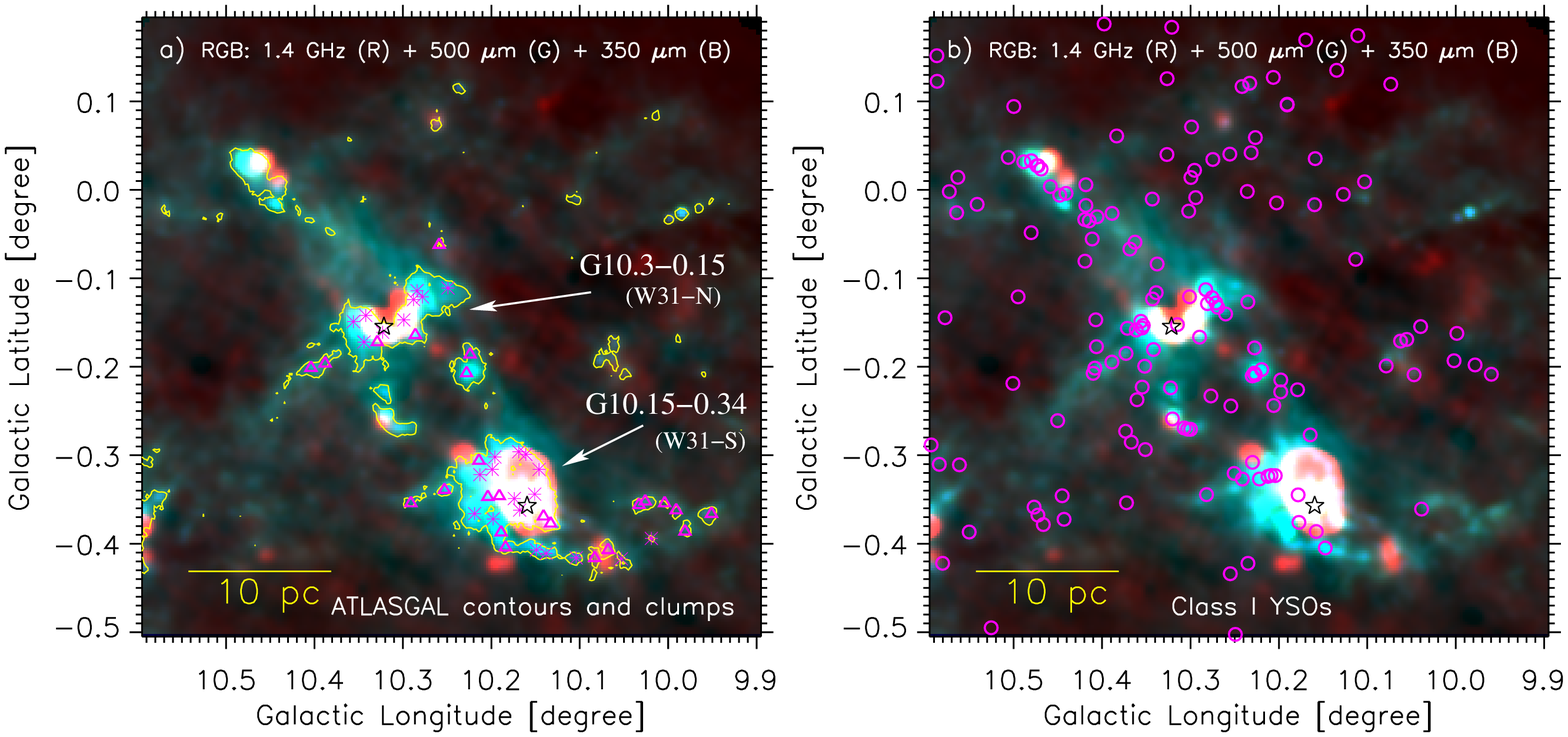}
\caption{Multi-wavelength view of an area hosting the W31 complex (size $\sim$0$\degr$.7 $\times$ 0$\degr$.695; centered at {\it l} = 10$\degr$.247; {\it b} = $-$0$\degr$.156). a) Overlay of the positions of the ATLASGAL 870 $\mu$m dust continuum clumps \citep[see triangles and asterisks;][]{Atlasgal} and the ATLASGAL 870 {\mic} dust continuum emission contour (at a level of 0.25 Jy beam$^{-1}$) on a three-color composite map. The color composite map is made of the NVSS 1.4 GHz (in red), {\it Herschel} 500 $\mu$m (in green), and {\it Herschel} 350 $\mu$m (in blue) images. All the ATLASGAL clumps are located at a distance of 3.55 kpc \citep{Atlasgal}. Asterisks highlight the ATLASGAL clumps associated with molecular outflows \citep[see][for more details]{Yang2022}. 
Two sites, G10.3$-$0.15 (or W31-N) and G10.15$-$0.34 (or W31-S), are indicated in the color composite map. 
b) Overlay of the positions of {\cone} (see circles) on the color-composite map as presented in Figure~\ref{fig1}a (see Section~\ref{dis_alt} for more details). 
In each panel, stars indicate the NVSS radio continuum peaks. 
A scale bar corresponding to 10 pc (at d = 3.55 kpc) is drawn in each panel.}
\label{fig1}
\end{figure*}

To infer the locations of the ionized emission and the dust continuum emission, we have produced a three-color composite map, which is composed of the NVSS 1.4 GHz (in red), {\it Herschel} 500 $\mu$m (in green), and {\it Herschel} 350 $\mu$m (in blue) images (see Figures~\ref{fig1}a and~\ref{fig1}b). 
The {\it Herschel} maps show extended structures of dust emission toward W31-N and W31-S, and each extended structure surrounds the ionized emission traced in the NVSS 1.4 GHz radio continuum map.  In Figure~\ref{fig1}a, we display an overlay of the ATLASGAL 870 {\mic} dust continuum emission contour (in yellow) on the color composite map, which is also overlaid 
with the positions of 49 ATLASGAL clumps \citep[from][]{Atlasgal}.  All these clumps are depicted in a velocity range of [8.9, 15.3] {\kps} and are located at a distance of 3.55 kpc \citep[see][for more details]{Atlasgal}.  The distribution of the ATLASGAL clumps enables us to infer a single extended physical system hosting W31-N and W31-S.  The southern area is seen with a large number of ATLASGAL clumps compared to the northern area.  Among the 49 ATLASGAL clumps, 25 clumps are associated with molecular outflows (see asterisks in Figure~\ref{fig1}a). \citet{Yang2022} tabulated the velocity ranges of the $^{13}$CO(J = 2--1) blue wing-like component
and red wing-like component for the ATLASGAL clumps associated with the molecular outflow.  In this connection, we find 9 and 16 clumps associated with outflows toward the northern region (W31-N) and the southern region (W31-S), respectively.  In general, the detection of a molecular outflow is known as a reliable signature of star formation activity.  Hence, star formation activities (including massive star formation) are traced toward both the regions W31-N and W31-S. 

The knowledge of infrared-excess sources or young stellar objects (YSOs) or embedded protostars is also used to depict star formation activity in a given star-forming region.  In general, one can select YSOs through the excess in the infrared emission originating from their dusty circumstellar disks and envelopes. To examine the locations of younger protostars in W31, we identified Class~I YSOs using the {\it Spitzer} 3.6--5.8 {\micro} photometric data of point-like sources.  The {\cone} are selected using the color conditions $[4.5] $-$ [5.8]\geq 0.7$ and $[3.6] $-$ [4.5]\geq 0.7$ \citep[see][for more details]{hartmann_2005,getman_2007}. Earlier, \citet{evans_2009} reported a mean age of Class~I YSOs to be $\sim$0.44 Myr.  In Figure~\ref{fig1}b, the positions of the detected Class~I YSOs are overlaid on the color composite image as shown in Figure~\ref{fig1}a, exhibiting the presence of protostars toward W31-N and W31-S. We find relatively more number of Class~I YSOs toward W31-N compared to W31-S.  

In Figure~\ref{fig1}a the {\it Herschel} and ATLASGAL dust continuum maps hint at the presence of elongated filamentary features toward W31-N and W31-S. These features are seen with higher {H$_2$} column density values of $\sim$10$^{22}$ -- 10$^{23}$ cm$^{-2}$ in the {\it Herschel} column density map (see Figure~\ref{fig2}a). In Figure~\ref{fig2}b, the {\it Herschel} temperature map is presented, and the site W31-S is prominently extended compared to the site W31-N. Both the H\,{\sc ii} regions are seen with the warm dust emission (i.e., $T_\mathrm{d}$ $\sim$21--32~K), while the elongated filamentary features seem to be associated with the relatively cold dust emission (i.e., $T_\mathrm{d}$ $\sim$17--20~K). It seems that O-type stars heat the dust in W31-N and W31-S.
%
%
%
%
%

\subsection{Hub-filament systems in W31}
\label{hfss}
\begin{figure*}
\epsscale{1}
\plotone{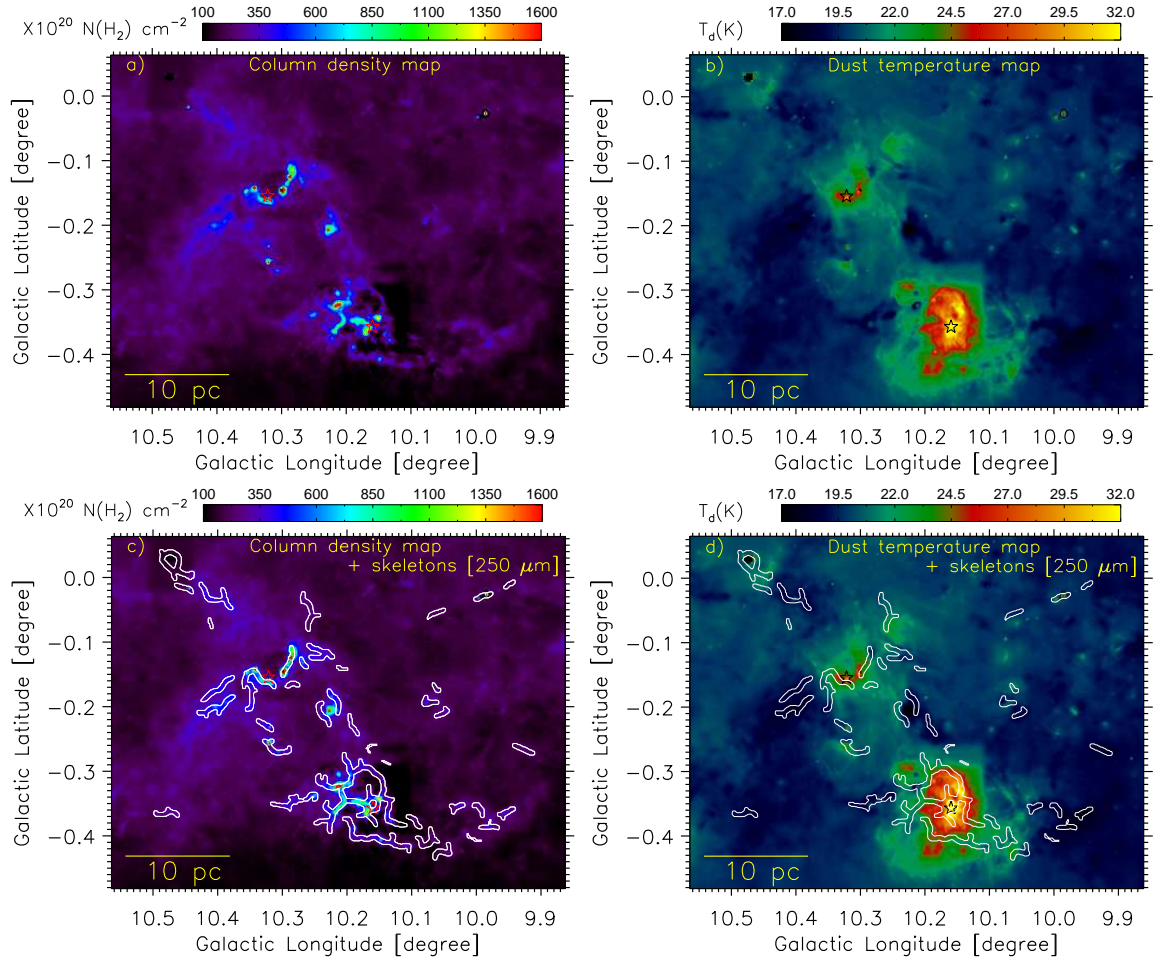}
\caption{a) The {\it Herschel} column density and b) dust temperature map of the W31 complex. The filamentary skeletons are highlighted in c) the {\it Herschel} column density map and d) the {\it Herschel} temperature map. The skeletons are produced from the {\it Herschel} 250{\mic} image using the alogorithm ``getsf" \citep{getsf_2022}. 
In each panel, stars are the same as shown in Figure~\ref{fig1}a.}
\label{fig2}
\end{figure*}

In general, the search for an HFS in a massive star-forming region hosting several O-type stars is a difficult task because of the intense energetic feedback of O-type stars (i.e., stellar wind, ionized emission, and radiation pressure) may affect their surroundings.  A similar situation is applicable for W31 containing several O-type stars, which has the complex structure as traced in the multi-wavelength images (see Figures~\ref{fig1}a and~\ref{fig2}a).

To identify the structures of filaments or filamentary skeletons in W31, we employed an algorithm ``getsf" \citep{getsf_2022} on the {\it Herschel} 250 {\micro} image (resolution $\sim$12$''$). 
The algorithm disintegrates the astronomical image into its structural components (i.e., sources and filaments) and splits these components from each other and their backgrounds.
One can find more details of the algorithm in \citet{getsf_2022} \citep[see also][]{bhadari22}. The algorithm needs a single user-deﬁned input, which is the maximum size of the structure to be extracted (for both filaments and sources).  Here, the ﬁlamentary skeletons in the {\it Herschel} 250 {\micro} are extracted with maximum source size of 40$''$ and maximum filamentary size of 220$''$.
%
%
%
In the direction of both the sites W31-S and W31-N, several parsec-scale filaments are identified and are highlighted in the {\it {Herschel}} column density and temperature maps, respectively (see Figures~\ref{fig2}c and~\ref{fig2}d). Filaments seem to be directed toward the W31-N and W31-S {\htwo} regions, which are associated with higher column density values and warm dust emission.  Such configurations can be considered as HFSs seen toward each {\htwo} region powered by several O-type stars in W31.
%
%
%
\subsection{Distribution of molecular gas in W31}
\label{sec:allmol}
%
\subsubsection{Spatial and velocity structures of molecular cloud associated with W31}
\label{sec:NANTEN}
\begin{figure*}
\epsscale{1}
\plotone{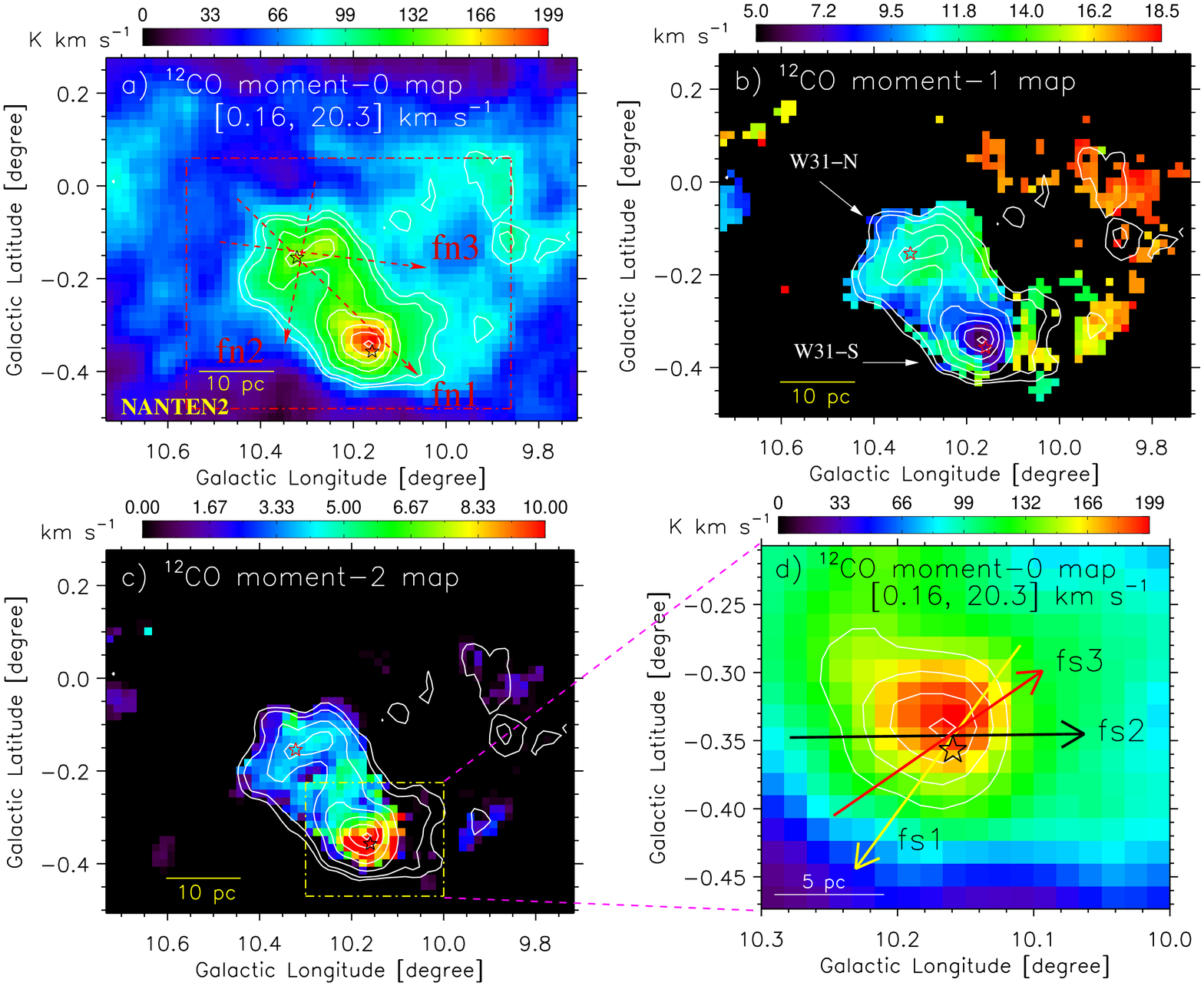}
\epsscale{1}
\plotone{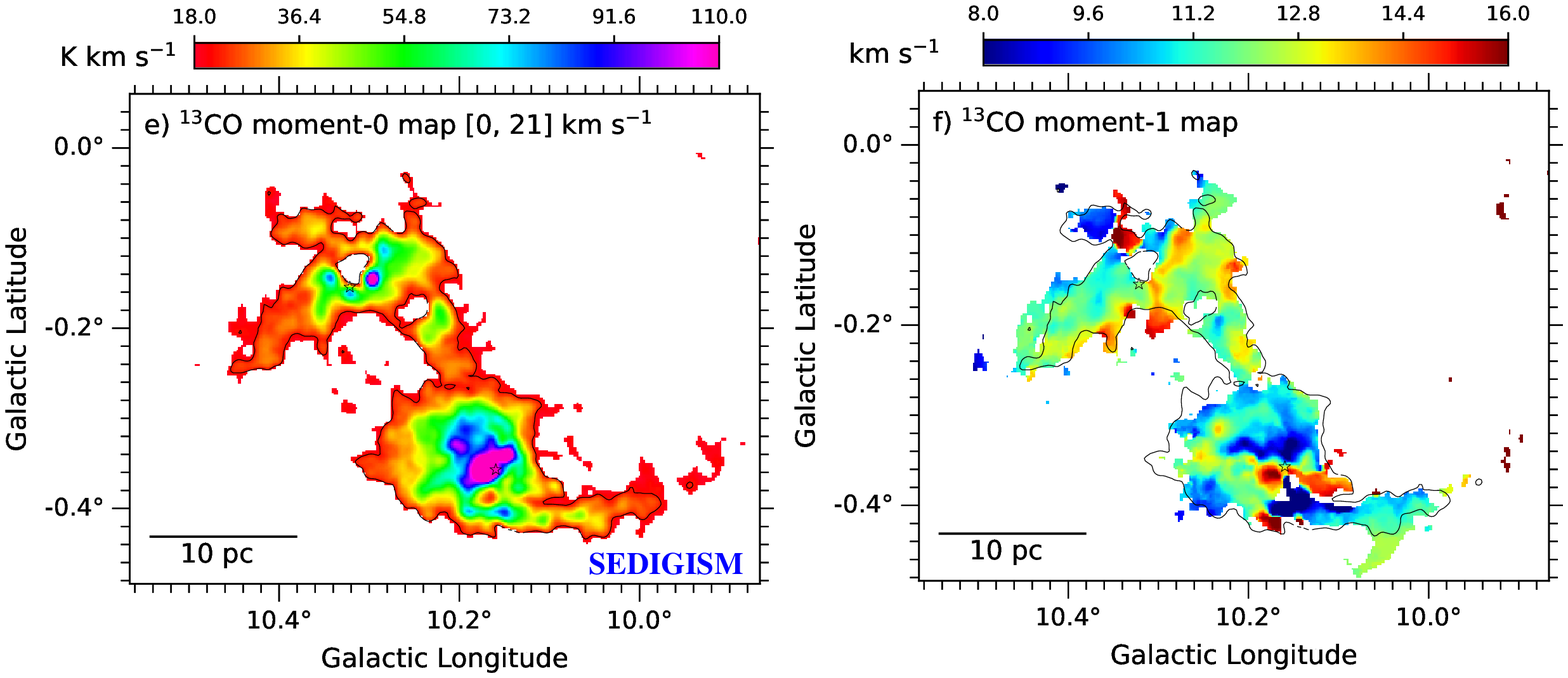}
\caption{a) The panel shows the NANTEN2 $^{12}$CO(J=1--0) integrated intensity (moment-0) map (at [0.16, 20.3] {\kps}) of an area hosting the W31 complex. 
A dotted dashed box (in red) highlights the area shown in Figure~\ref{fig2}a.  The molecular map is also overlaid with the molecular emission (in contours). 
Three arrows ``fn1", ``fn2", and ``fn3" are shown in the panel, where the position-velocity diagrams are produced (see Figures~\ref{fig4}a--\ref{fig4}c). 
b) Overlay of the molecular emission contours on the NANTEN2 {\twco} moment-1 map.  c) Overlay of the molecular emission contours on the NANTEN2 {\twco} moment-2 map. 
A dotted-dashed box (in yellow) represents an area presented in Figure~\ref{fig3}d. In panels ``a--c", the molecular integrated emission contours are shown with the levels of 45, 50, 60, 70, 80, 90, and 98\% of the peak value (i.e., 198.7 K {\kps}). d) A zoomed-in view of the NANTEN2 moment-0 map at [0.16, 20.3] {\kps} toward W31-S (see a dotted dashed box in Figure~\ref{fig3}c). The levels of the contours are 70, 80, 90, and 98\% of the peak integrated emission value (i.e., 198.7 K {\kps}). Three arrows ``fs1", ``fs2", and ``fs3" are marked in the panel, where the position-velocity diagrams are generated (see Figures~\ref{fig4}d--\ref{fig4}f). e) The panel displays the SEDIGISM {\tco} moment-0 map at [0, 21] {\kps} toward the W31 complex hosting both W31-N and W31-S. 
The {\tco} emission contour at 22.0 K {\kps} is also overplotted on the SEDIGISM moment-0 map.
f) The SEDIGISM moment-1 map of the {\tco} emission toward the W31 complex.  The {\tco} emission contour, as shown in Figure~\ref{fig3}e, is also overplotted on the moment-1 map. 
In each panel, stars are the same as shown in Figure~\ref{fig1}a.}
\label{fig3}
\end{figure*}

%
%
We have examined the NANTEN2 $^{12}$CO(J = 1--0) emission and the SEDIGISM $^{13}$CO(J = 2--1) emission to infer the cloud morphology of the W31 complex. 
Figure~\ref{fig3}a presents an integrated intensity (moment-0) map of the NANTEN2 {\twco} emission for a wide region of the W31 complex, where the molecular gas is integrated over a velocity range of [0.16, 20.3] {\kps}.  A cloud structure extending from north-east to south-west is evident in the NANTEN2 moment-0 map, and molecular peaks nearly coincide with the NVSS radio continuum peaks observed toward W31-N and W31-S (see stars in Figure~\ref{fig3}a).  Figure~\ref{fig3}b displays the moment-1 map of the NANTEN2 $^{12}$CO emission, enabling us to examine the velocity distribution of the gas.  We find a noticeable velocity gradient toward the cloud based on the velocities observed toward W31-N and W31-S.  The moment-2 map of the NANTEN2 $^{12}$CO emission is shown in Figure \ref{fig3}c, exhibiting higher velocity dispersion values ($>$ 5 {\kps}) toward W31-N and W31-S.  A zoomed-in view of the molecular cloud associated with W31-S is presented in Figure~\ref{fig3}d (see a dotted dashed box in Figure~\ref{fig3}c). 

The NANTEN2 $^{12}$CO line data reveal the overall distribution and kinematics of the molecular cloud associated with the W31 complex, but the NANTEN2 line data are limited with the resolution.  The SEDIGISM $^{13}$CO(J = 2--1) line data (beam size $\sim$41{\as}.4) have a better resolution compared the NANTEN2 $^{12}$CO(J = 1--0) emission. 
The SEDIGISM $^{13}$CO(J = 2--1) integrated line intensity map at [0, 21] {\kps} is presented in Figure~\ref{fig3}e. Hence, in Figure~\ref{fig3}e, we obtain more insights into the cloud morphology associated with the W31 complex, allowing us to examine the molecular structures observed toward W31-N and W31-S. 
In Figure~\ref{fig3}f, the moment-1 map of the SEDIGISM {\tco} is displayed, revealing the presence of a noticeable velocity gradient toward W31-N and W31-S.
%

\begin{figure*}
\epsscale{1}
\plotone{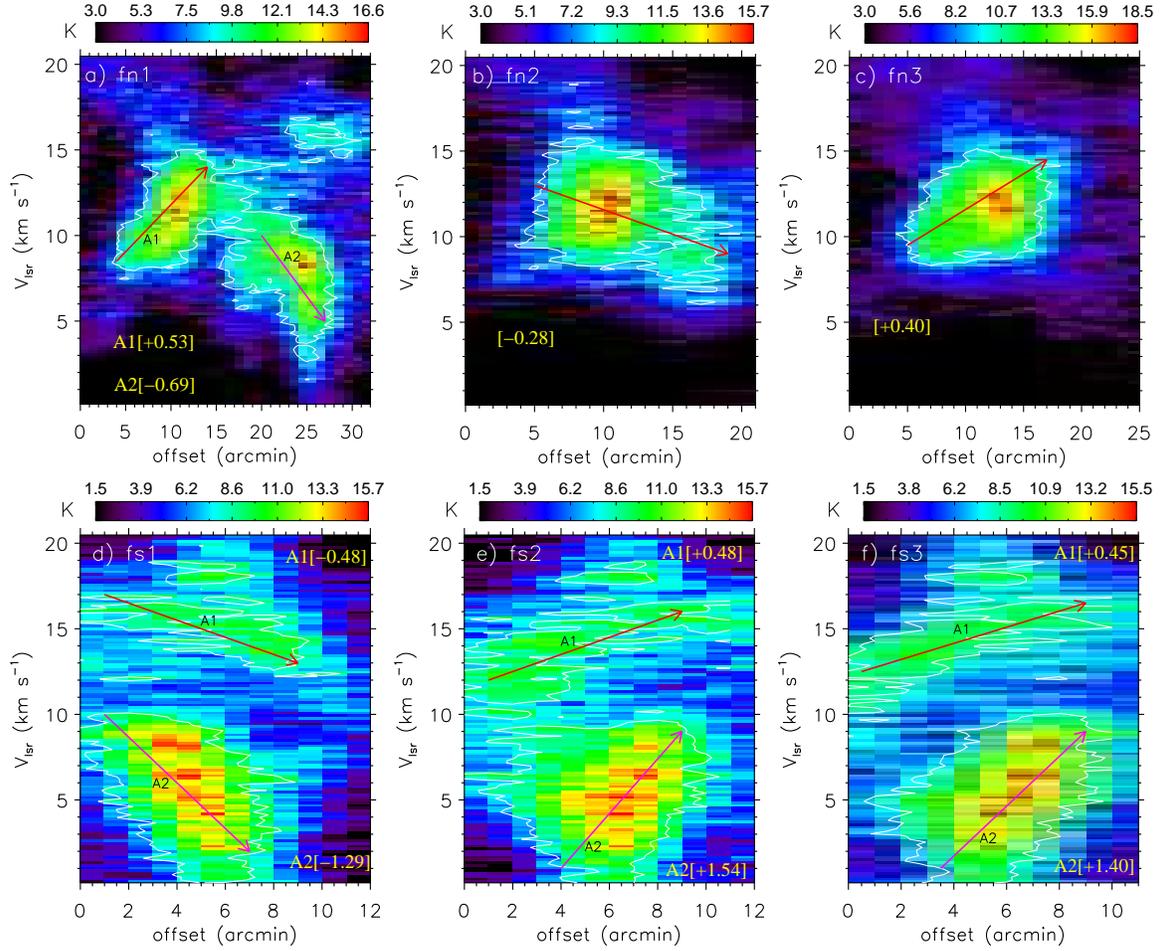}
\caption{Position-velocity diagrams of the {\twco} emission along arrows a) ``fn1"; b) ``fn2"; c) ``fn3"; d) ``fs1"; e) ``fs2"; f) ``fs3")  (see arrows in Figures~\ref{fig2}a and~\ref{fig2}d). In each panel, a contour is shown with a level of 53\% of its corresponding peak value (in K). In all panels, a velocity gradient is computed along the arrow(s), and is mentioned in the unit of {\kps}pc$^{-1}$.}
%
\label{fig4}
\end{figure*}

We have produced several position-velocity diagrams to study velocity structures of the molecular cloud associated with W31.  
In Figures~\ref{fig4}a,~\ref{fig4}b, and~\ref{fig4}c, using the NANTEN2 $^{12}$CO(J = 1--0) line data, we present position-velocity diagrams along the arrows ``fn1", ``fn2", and ``fn3" as indicated in Figure~\ref{fig3}a, respectively. The width of the slices used to extract position-velocity diagrams is 1 pixel (= 60$''$).
The arrow "fn1" passes through both the sites W31-N and W31-S, while the other two arrows are selected toward W31-N.  Figures~\ref{fig4}d,~\ref{fig4}e, and~\ref{fig4}f show position-velocity diagrams along the arrows ``fs1", ``fs2", and ``fs3" as marked in Figure~\ref{fig3}d, respectively. In each panel of Figure~\ref{fig4}, one or two arrows are drawn to show the associated velocity gradients.  The velocity gradients measured along the arrows shown in the panels a), b) and c) are [$+$0.53 (for A1) and $-$0.69 (for A2)], $-$0.28 and  $+$0.40 {\kps}pc$^{-1}$, respectively. Similarly, in the last three panels d), e) and f) velocity gradient values along the arrows are estimated to be [$-$0.48 (A1) and $-$1.29 (A2)], [$+$0.48 (A1) and $+$1.54 (A2)] and [$+$0.45 (A1) and $+$1.40 (A2)] {\kps}pc$^{-1}$, respectively. 
Note that several ATLASGAL clumps associated with molecular outflows and young protostars are detected toward W31-N and W31-S.  Hence, star formation activities may be one of the factors to explain the observed velocity gradients.  Furthermore, the observed velocity gradients could also suggest the existence of multiple velocity components. 
In the direction of W31-S, the NANTEN2 $^{12}$CO line data hint at the presence of two velocity components and higher values of velocity dispersion. 
Such observed results are further investigated using the SEDIGISM {\tco} line data in Section~\ref{sec:SEDIGISM}. 
However, we do not find a clear signature of the presence of multiple velocity components toward W31-N (see also section~\ref{sec:4.2}).
%
%
%
\subsubsection{Velocity components toward W31-S}
\label{sec:SEDIGISM}
\begin{figure*}
\epsscale{0.57}
\plotone{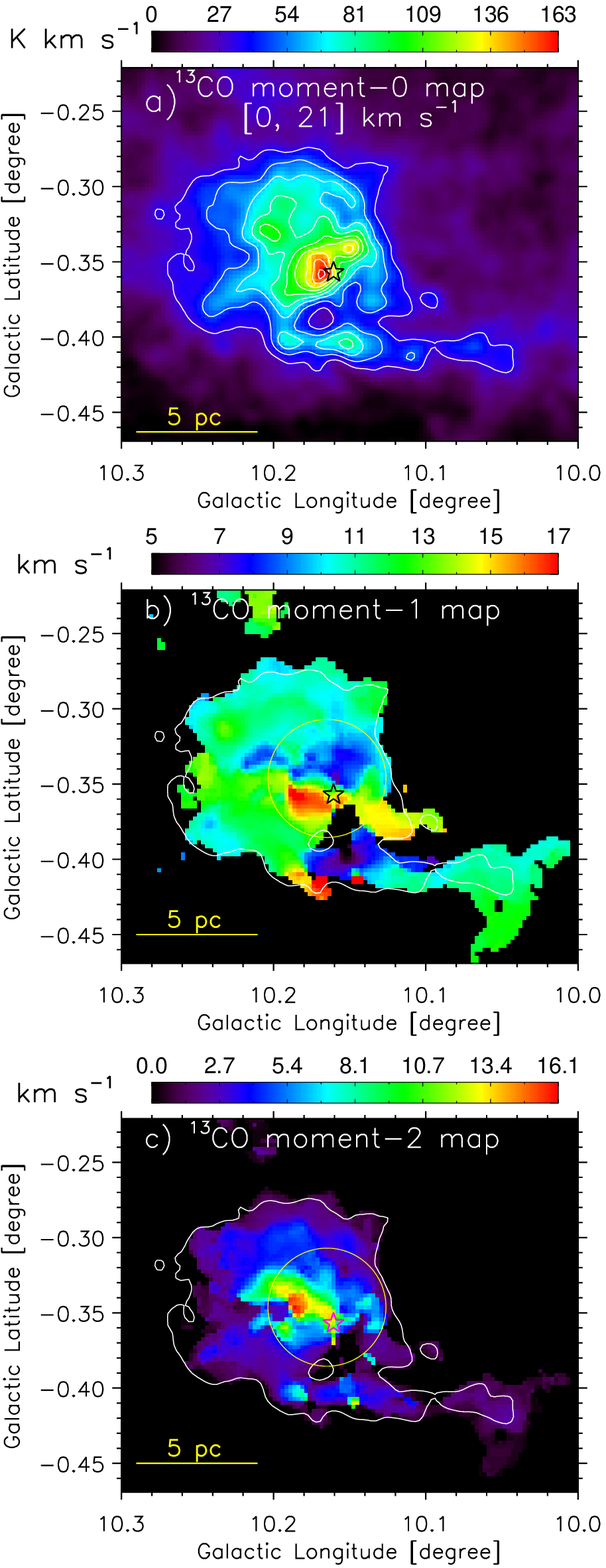}
\caption{A zoomed-in view of moment maps of the SEDIGISM {\tco} emission toward W31-S (see a dotted dashed box in Figure~\ref{fig3}c).
a) {\tco} moment-0 map at [0, 21] {\kps}. b) {\tco} moment-1 map. c) {\tco} moment-2 map. In each moment map, the {\tco} emission contour is also shown with a level of 32.6 K {\kps}. A circle is indicated in the moment-1 and moment-2 maps to locate the area where the NVSS 1.4 GHz emission is mainly distributed toward the site W31-S. 
A star is the same as marked in Figure~\ref{fig3}d.}
\label{fig5}
\end{figure*}

%
In the direction of W31-S, Figures~\ref{fig5}a,~\ref{fig5}b, and~\ref{fig5}c display 
the moment-0, moment-1, and moment-2 maps of the SEDIGISM $^{13}$CO(J = 2--1) emission, respectively (see a dotted-dashed box in Figure~\ref{fig3}c). In the moment-0 map, the molecular gas is integrated from 0 to 21 {\kps}.  The contours of the moment-0 map and the NVSS radio continuum peak are highlighted in all the SEDIGISM moment maps.  The peak of the molecular emission coincides with the NVSS radio peak. 
The moment-1 map shows a very obvious velocity gradient toward an area highlighted 
by a circle in Figure~\ref{fig5}b, where the NVSS radio peak is detected. 
From the moment-1 map, we might see two different velocity components (around 8 and 16 {\kps}) 
and an intermediate velocity component (around 12 {\kps}) in the direction of the area highlighted by a circle. In the moment-2 map, higher values of velocity dispersion (i.e., 13--16 {\kps}) are found toward the area enclosed by a circle (see Figure~\ref{fig5}c) as observed in the NANTEN2 {\twco} moment-2 map.
The velocity dispersion value seems to decay monotonically toward the outer parts of the cloud associated with W31-S. 
%
 
\begin{figure*}
\epsscale{0.88}
\plotone{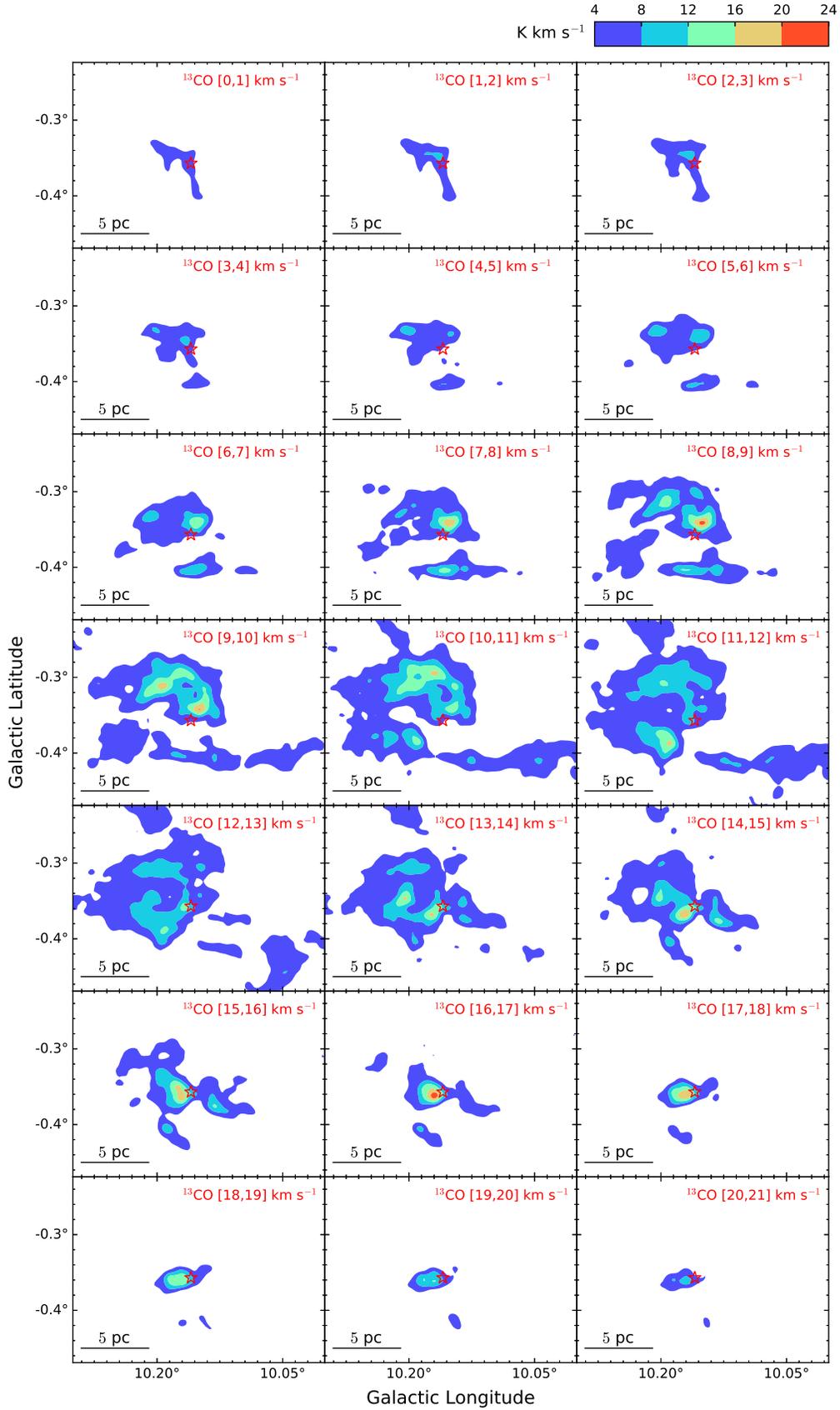}
\caption{Velocity channel maps of the SEDIGISM {\tco} emission with a velocity interval of 1 {\kps} from 0 to 21 {\kps} toward W31-S. 
A color bar is shown to the top right corner. 
A star is the same as marked in Figure~\ref{fig3}d.}
\label{fig6}
\end{figure*}

The velocity channel maps of $^{13}$CO(J = 2--1) are shown in Figure \ref{fig6}. 
A total of 21 panels are generated, starting from 0 to 21 {\kps} with an integration interval of 1 {\kps}.  In Figure \ref{fig6}, the NVSS radio peak is also marked in each panel.  The channel maps seem to support the proposal of the presence of two velocity components toward W31-S (see panels at [10, 11] and [14, 15] {\kps}).

\begin{figure*}
\epsscale{1}
\plotone{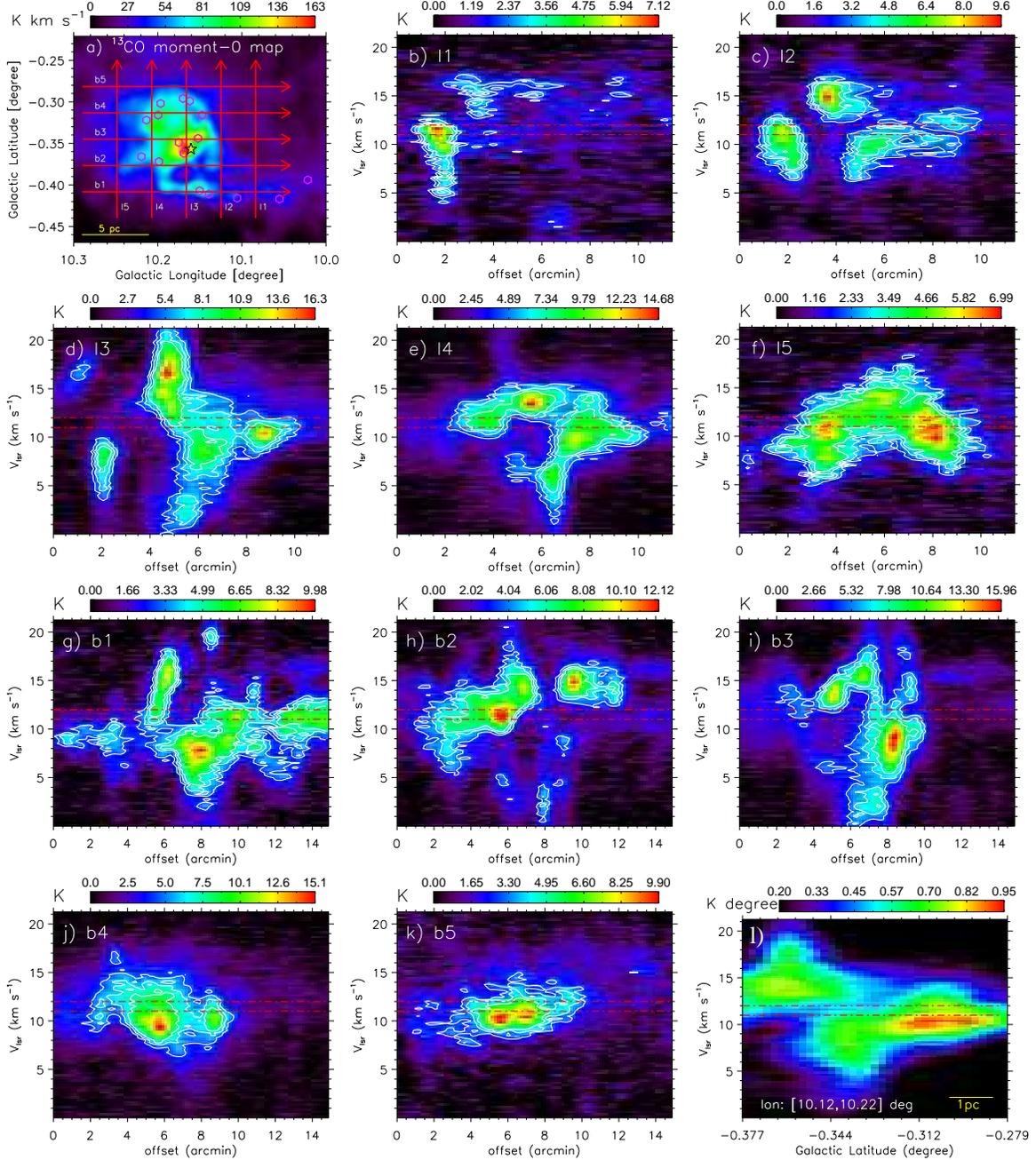}
\caption{a) The panel presents the moment-0 map of the SEDIGISM {\tco} emission in the direction of W31-S 
(see Figure~\ref{fig5}a). 
A star is the same as marked in Figure~\ref{fig3}d.  Hexagons (in magenta) indicate the positions of the ATLASGAL clumps associated with the molecular outflows 
(see also asterisks in Figure~\ref{fig1}a), and velocity ranges of wing components of these outflows are presented in Figures~\ref{fig8}a and~\ref{fig8}b. Ten arrows (5 vertical: ``l1--l5"; 5 horizontal: ``b1--b5") are also indicated in the panel, where the position-velocity diagrams are produced (see Figures~\ref{fig7}b--\ref{fig7}k). Position-velocity diagrams along the vertical arrows b) ``l1''; c) ``l2''; d) ``l3'', e) ``l4''; and f) ``l5'', respectively as shown in Figure~\ref{fig7}a. Position-velocity diagrams along the horizontal arrows g) ``b1''; h) ``b2''; i) ``b3'', j) ``b4''; and k) ``b5'', respectively as shown in Figure~\ref{fig7}a. l) The panel shows the latitude-velocity diagram of the SEDIGISM {\tco} emission. The longitude range used in the integration is indicated in the panel. 
In panels ``b--k", the position-velocity diagrams are overplotted with the contours with the levels of 30, 40 and 50\% of corresponding peak values. In each position-velocity diagram, two dotted dashed lines (in red) at V$_\mathrm{lsr}$ = 11 and 12 {\kps} are marked.}
\label{fig7}
\end{figure*}

\begin{figure*}
\epsscale{0.85}
\plotone{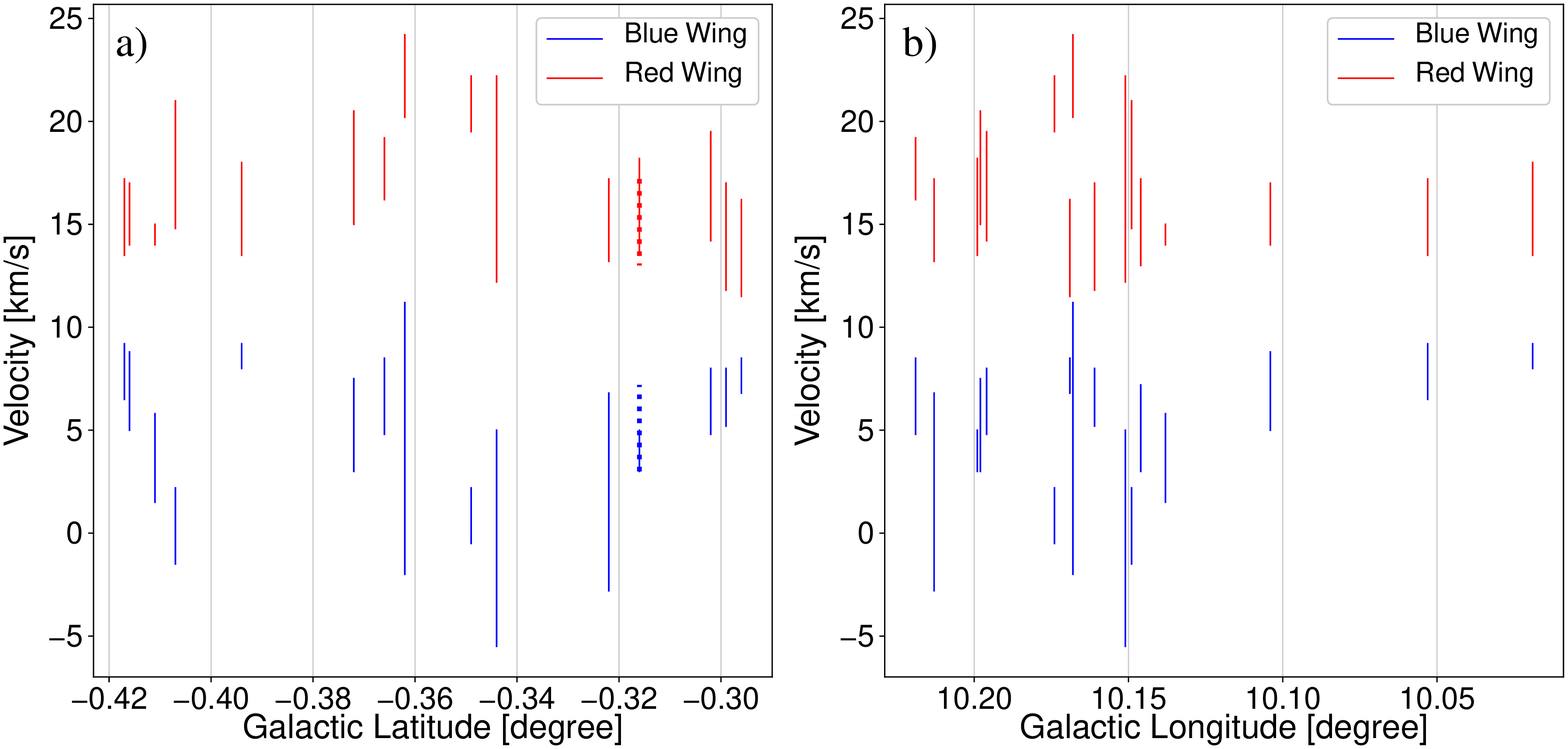}
\epsscale{1}
\plotone{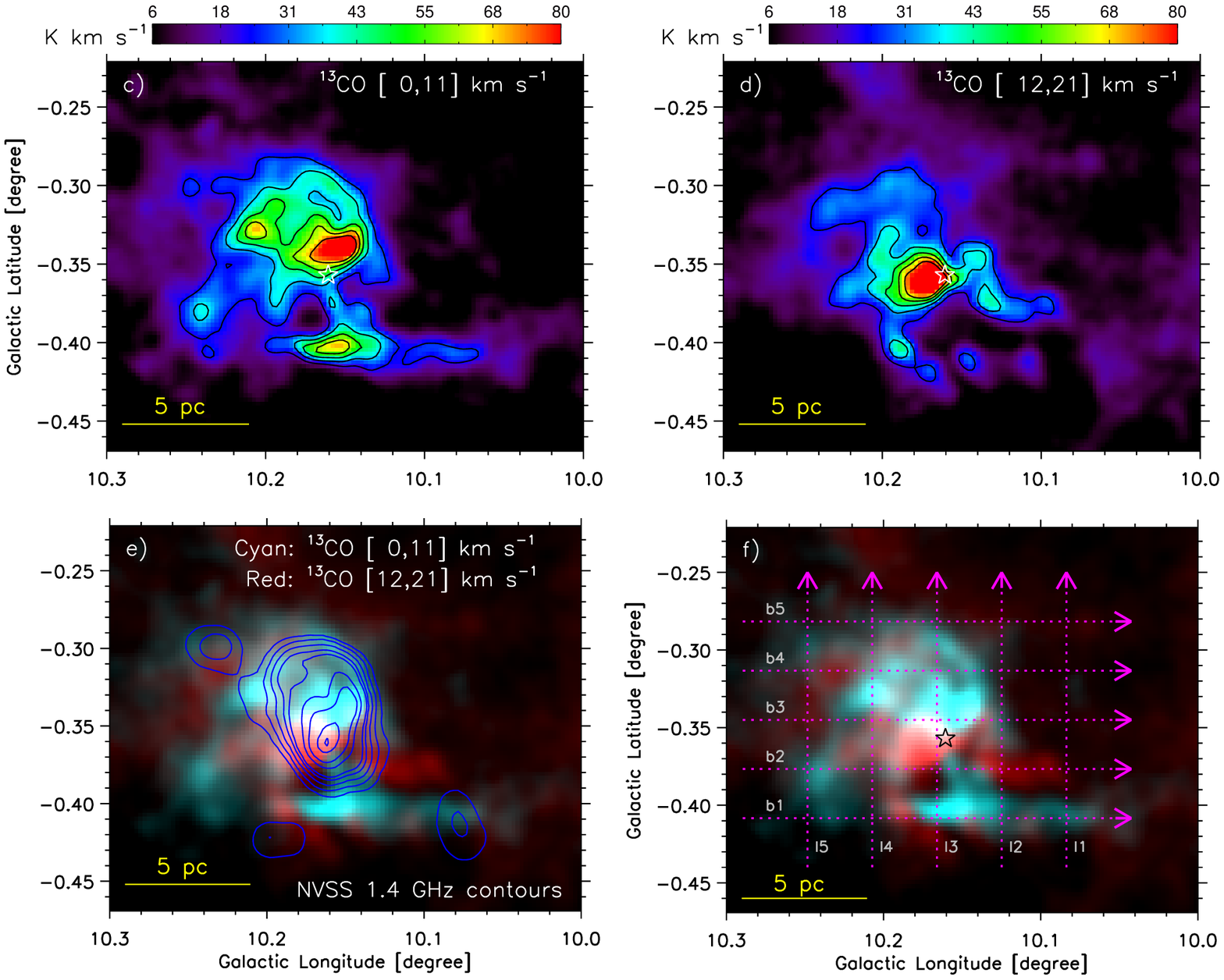}
\caption{a) Velocity ranges of the $^{13}$CO(J = 2--1) blue wing-like component and red wing-like component of molecular outflows 
associated with ATLASGAL clumps are shown against their latitudes (two clumps sharing the same latitude -0$\degr$.316 are differentiated with solid and dotted lines). These clumps are distributed toward W31-S (see hexagons in Figure~\ref{fig7}a).  
b) Same as panel ``a,'' but velocity ranges are presented against longitudes of the clumps distributed toward W31-S (see hexagons in Figure~\ref{fig7}a).
c) Moment-0 map at [0, 11] {\kps} for the SEDIGISM {\tco} emission toward W31-S. 
d) Moment-0 map at [12, 21] {\kps} for the {\tco} emission toward the same region. 
In panels ``a" and ``b", the SEDIGISM {\tco} emission contours are at 24, 36, 48, 60, and 72 K {\kps}. 
e) The panel displays a two-color composite image made, using the SEDIGISM {\tco} map at [12, 21] {\kps} (in red) and the SEDIGISM {\tco} map at [0, 11] {\kps} (in cyan), overlaid with the NVSS radio continuum contours having the levels of 0.02, 0.08, 0.2, 0.4, 0.6, 1.4, 2.2, 3.7 and 4.4 Jy beam$^{-1}$ (here 1$\sigma$ $\sim$ 0.45 mJy beam$^{-1}$). f) The panel is the same as Figure~\ref{fig8}c, but overlaid with the several arrows as shown in Figure~\ref{fig7}a. In panels ``a", ``b", and ``d", a star is the same as marked in Figure~\ref{fig3}d.}
\label{fig8}
\end{figure*}

%
%
Figure~\ref{fig7}a represents the {\tco} moment-0 map of W31-S, where we have marked ten arrows (i.e., 5 vertical: ``l1--l5"; 5 horizontal: ``b1--b5"), where the position-velocity diagrams are extracted. The width of the slices used to extract position-velocity diagrams is 1 pixel (= 9\rlap.{$''$}5). Additionally, hexagons (in magenta) overplotted on Figure~\ref{fig7}a indicates the positions of clumps associated with outflows (see also asterisks in Figure~\ref{fig1}a). In Figures~\ref{fig7}b--\ref{fig7}k, we present the position-velocity diagrams for {\tco} emission along the ten arrows, respectively. Figure~\ref{fig7}l also displays the latitude-velocity diagram for {\tco} emission, where the molecular emission is integrated in the longitude range of [10.12 deg, 10.22 deg]. We have also examined the extents of the blue wing-like velocity component and the red wing-like velocity component of a molecular outflow associated with ATLASGAL clumps seen toward the W31-S region (see Figures~\ref{fig8}a and~\ref{fig8}b). Figures~\ref{fig8}a and~\ref{fig8}b exhibit the velocity ranges of outflow wings against latitude and longitude of each ATLASGAL clump distributed toward W31-S, respectively (see hexagons in Figure 7a).  We find that the knowledge of the velocity ranges of outflow wings is useful for exploring the complex velocity structure observed in the position-velocity diagrams (see Figures~\ref{fig7}b--\ref{fig7}l). 
%
%
In Figures~\ref{fig7}c,~\ref{fig7}d,~\ref{fig7}e, and~\ref{fig7}i, we suspect the presence of two velocity components, which are found in the direction of arrows ``l2", ``l3", ``l4", and ``b3", respectively. These arrows pass through areas with higher velocity dispersions and the zones associated with the radio continuum emission in W31-S.  Considering the locations of these arrows, Figure~\ref{fig7}l is produced to study the velocity structure, allowing us to observe a connection of two velocity components toward W31-S.  No spike or outflow signature is seen in Figure~\ref{fig7}l.  In the panel ``i'' (represents the arrow ``b3''), we may suggest both the presence of a spike (or outflow signature) and two velocity components.  Hence, our analysis suggests that there exist two velocity components along with outflow signatures toward W31-S.   

%

On the basis of the position-velocity diagrams, we produce two integrated maps of {\tco} at two different velocity ranges (i.e., [0, 11] and [12, 21] {\kps}). To examine the morphology of each cloud component, the integrated intensity maps and contours at [0, 11] and [12, 21] {\kps} are presented in Figures~\ref{fig8}c and~\ref{fig8}d, respectively.  In Figure~\ref{fig8}e, we present a two color-composite image generated using the {\tco} maps at [0, 11] and [12, 21] {\kps} in cyan and red, respectively, which is also overlaid with the NVSS radio continuum emission contours. The ionized emission is mainly concentrated toward the overlapping zones of these two cloud components. Figure~\ref{fig8}f is the same as Figure~\ref{fig8}e, but identical arrows from the Figure~\ref{fig7}a are highlighted, enabling us to assess the presence of two cloud components in the direction of the arrows. 

\begin{figure*}
\epsscale{1}
\plotone{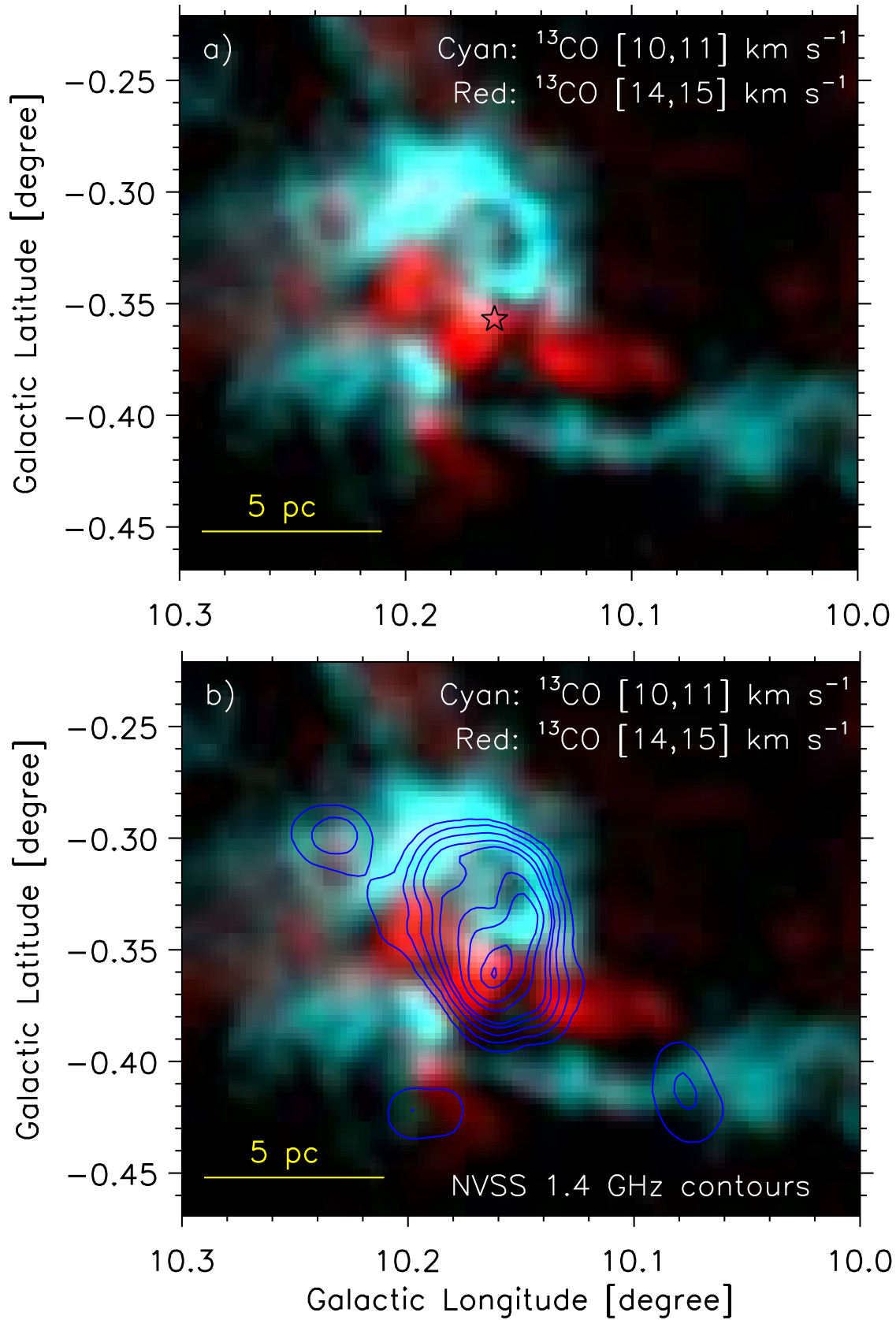}
\caption{a) The panel presents a two-color composite image made using the SEDIGISM {\tco} map at [14, 15] {\kps} (in red) and the SEDIGISM {\tco} map at [10, 11] {\kps} (in cyan). A star is the same as marked in Figure~\ref{fig3}d. b) The panel is the same as Figure~\ref{fig9}a, but overlaid with the NVSS radio continuum contours as mentioned in Figure~\ref{fig8}c.}
\label{fig9}
\end{figure*}

From the channel maps of {\tco}, we find an ``intensity-enhancement" feature in the map at [14, 15] {\kps} and an ``intensity-depression" feature in the map at [10, 11] {\kps}. Figure~\ref{fig9}a shows a two color-composite image using the {\tco} maps at [10, 11] and [14, 15] {\kps} in cyan and red, respectively.
Figure~\ref{fig9}b is the same as Figure~\ref{fig9}a but is superimposed with the NVSS radio continuum emission contours. 
In Figure~\ref{fig9}a, we find a spatial match of the ``intensity-enhancement" and ``intensity-depression" features, which can be referred to as a complementary distribution of clouds.  Figure~\ref{fig9}b displays the presence of the NVSS radio continuum emission toward the complementary distribution of clouds.  In this relation, the ionized emission seems to be distributed toward the ``intensity-depression" feature.

The implication of all these findings is discussed in Section~\ref{sec:disc}.
\section{Discussion}
\label{sec:disc}
\subsection{W31-S: A site of CCC}
\label{subsec:sfc}
Based on the distribution of molecular gas and dust clumps at a distance of 3.55 kpc, this work confirms the existence of a single extended complex W31 containing two {\htwo} regions, W31-N and W31-S, which are known to be excited by a cluster of O-type stars.  For the first time, with the help of the {\it Herschel} 250 {\micro} image, we propose the existence of a HFS toward both W31-N and W31-S (see Section~\ref{dis_alt}). In the HFSs, massive stars form in the densest part of the clouds (i.e., the central hub), where several filaments converge.  Filaments are thought to channel interstellar gas and dust to the central hub that eventually forms low- and high-mass stars \citep{myers_2009,andre_2010,schenider_2012,motte_2018}.  In recent years, researchers often employ the clump-fed scenario to explain the birth of massive stars in HFSs (see Section~\ref{sec:intro}).  However, the existence of HFSs is still hotly debated.

Interestingly, the inspection of the NANTEN2 $^{12}$CO(J = 1--0) emission has provided new insight into the study of molecular cloud associated with W31, which is the presence of two velocity components toward W31-S (see Section~\ref{sec:NANTEN}). 
This outcome is further examined using the SEDIGISM $^{13}$CO(J = 2--1) line data, which allowed us to carefully study 
the kinematics associated with the entire cloud.
In the position-velocity maps, it is difficult to trace an individual clump's outflow kinematics.  Still, one can see the kinematics of the entire cloud  (see Section~\ref{sec:SEDIGISM}).  In the direction of W31-S, the SEDIGISM $^{13}$CO(J = 2--1) line data confirm the presence of two velocity components around 8 and 16 \kps, which are also connected in velocity space.  We also suggest the presence of outflow signatures toward W31-S.
Additionally, in the {\tco} maps, a complementary distribution, which is a spatial fit of ``key/intensity-enhancement" and ``cavity/keyhole/intensity-depression" features, is investigated toward W31-S (see Section~\ref{sec:SEDIGISM}).

In a collision site, a spatial and velocity connection of two cloud components is considered as a reliable characteristic of CCC \citep[e.g.,][]{torri_2011,torri_2015,torri_2017,torri_2021,fukui_2014,fukui_2015,CCC3,fukuia_2018,fukui21,dewangan_2017,nishimura_2017,hayashi_2018,sano_2017,sano_2018,fujita21}. Furthermore, two colliding clouds show a complementary distribution in position-position space \citep[e.g.,][]{fukui18,Dewangan2018,fujita21} and a bridge feature (a low intensity flattened profile between two velocity peaks) in velocity space \citep[e.g.,][]{haworth_2015,torri_2017,Dewangan2017,Priestley2021}.  
Here, a bridge feature represents a compressed layer of gas produced through the collision event \citep{haworth_2015,haworth_2015b}.
%
%
We also highlight some works on CCC concerning extragalactic objects 
(e.g., LMC R136 \citep{fukui17exg}, LMC N44 \citep{tsuge19exg}, Antennae galaxies \citep{tsuge21exg}, M33 \citep{sano21exg}). The head-on collision of the two clouds creates a cavity due to the conservation of linear momentum on the structure of the larger cloud. 

\begin{figure*}
\epsscale{1}
\plotone{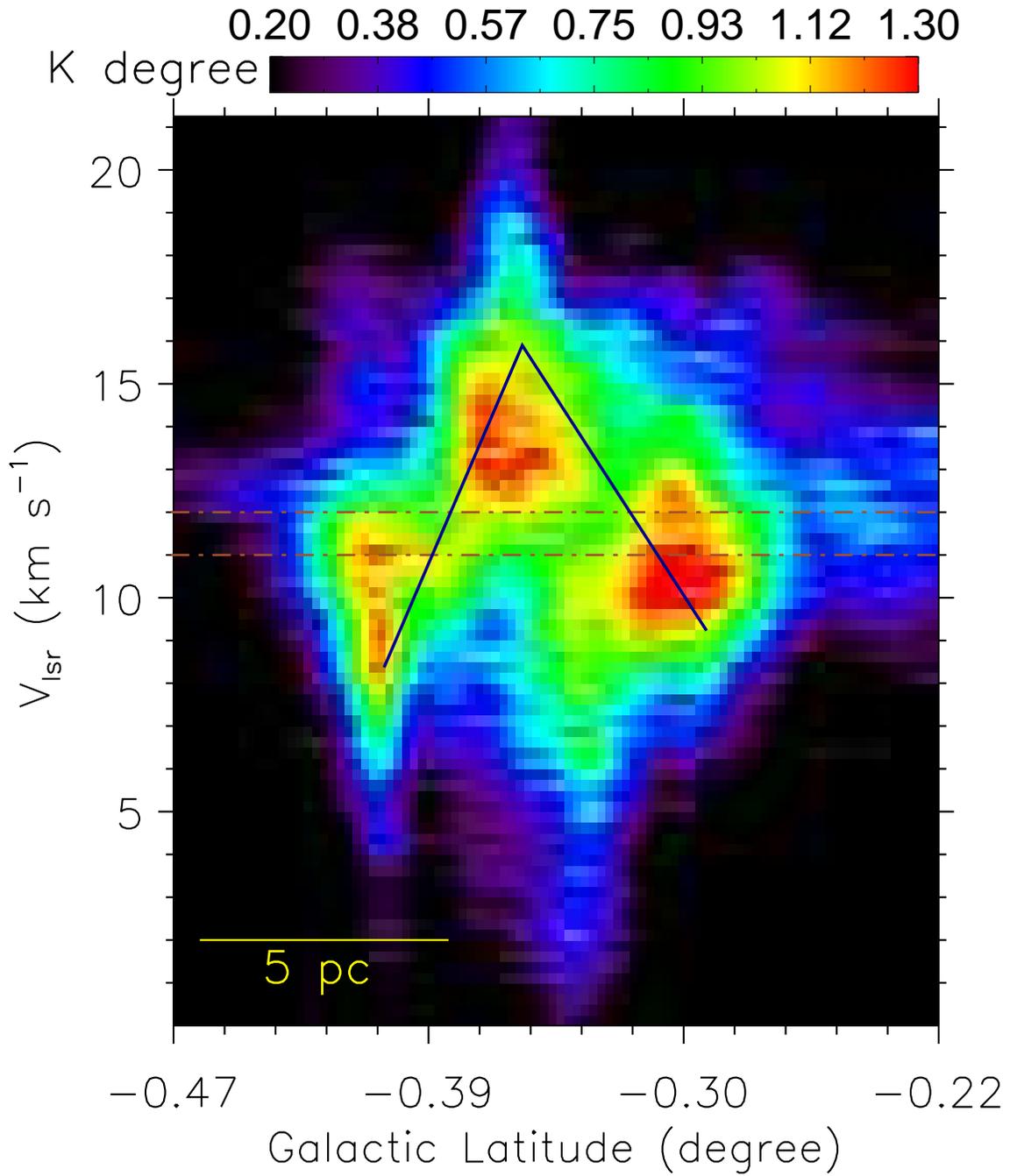}
\caption{Latitude-velocity diagram of the {\tco} emission. An inverted ``V" like structure is also indicated in the figure. 
Two dotted dashed lines at V$_\mathrm{lsr}$ = 11 and 12 {\kps} are marked in the figure.}
\label{fig10}
\end{figure*}

An observer whose viewing axis makes an acute angle to the axis of collision finds a complementary distribution between the smaller cloud and the structure of the cavity in the larger cloud \citep[e.g.,][]{takahira_2018,fukuia_2018,Enokiya2021,Nishimura2021}.
In the present case, no displacement between ``key/intensity-enhancement" and ``cavity/keyhole/intensity-depression" is found,
suggesting that the line of sight is approximately parallel to the collision axis.
Additionally, almost a V-like velocity structure is observed in the latitude-velocity diagram of {\tco} (see Figure~\ref{fig10}).
It is thought that the two signatures of CCC (i.e., the complementary distribution and bridge feature) together may be observed as a ``V-shaped" in position-velocity space \citep[e.g.,][]{hayashi_2018,fukuia_2018,fukui21,torri_2021}. For inclined viewing angles to the colliding axis, the V-shaped distribution is skewed \citep[][see Figure 5 therein]{fukuia_2018}.  In our analysis, the ``V-shaped distribution" is not skewed.
Therefore, it suggests that the inclination of our viewing angle to the colliding axis is very small.

 The {\htwo} regions powered by O-type stars \citep{Blum} and signatures of star formation activities are traced toward the colliding interface region of two cloud components (see Figures~\ref{fig8}e and~\ref{fig9}b).  Hence, a deep connection between the birth of the O-type stars with CCC is likely.  As previously discussed, the inclination of our viewing angle to the axis of collision is very small.  Hence, we assume this value to be {\simi}10{\degr} to derive an estimation of the collision time scale.  The size of the overlapping region is found to be $\sim$2.5 pc, and its value along the line of collision is $2.5\; \text{pc}/\text{sin(10{\degr})}\approx14.3\,\text{pc}$. 
%
%
Radial velocity difference between two clouds is {\simi}8 {\kps} and its value along the line of 
collision is $8\; \text{\kps}/\text{cos(10{\degr})}\approx 8.12\; \text{\kps}$. 
Therefore, the collision time scale is determined to be 14.3 pc/8.12 {\kps} = 1.7 Myr. 
To form an O-type star (mass = $30\,${\Msolar}), it takes nearly 10$^5$ yr with a mass accretion rate of {\simi}3 $\times$ 10$^{-4}${\Msolar} yr$^{-1}$ \citep{wolfire_1987,inoue_2013}. Such a value of mass accretion rate is thought to be enough to overcome the high stellar radiation pressure. 
Therefore, a collision event had happened well before the onset of star-formation activities in W31-S. 
It implies that the formation of massive O-type stars and Class~I YSOs in W31-S appears to be triggered by the CCC process as predicted by numerical simulations.
\subsection{A larger picture of CCC in W31 ?}
\label{sec:4.2}
Earlier, based on the presence of 6.7 GHz methanol masers in the boundary of CN 148 MIR bubble \citep{CN148}, \citet{Beuther} suggested that W31-N is relatively younger than W31-S. 
However, the previous results derived using the near-infrared spectroscopic observations of point-like sources \citep [e.g.,][]{Blum,bik_2005} and radio continuum data \citep{W31Kim2002} suggest almost an identical age for both the regions, which is around 1 Myr. Additionally, several dust clumps associated with outflows and Class~I YSOs (age {\simi}0.44 Myr) are also seen toward both the sites (see Figures~\ref{fig1}a and \ref{fig1}b, respectively). It implies that an earlier stage of star formation activities around a relatively evolved {\htwo} region is evident in both the sites. These results, therefore, hint at the common origin of two sites, W31-N and W31-S, and encourage exploring the detailed gas kinematics toward the entire W31 complex.
%
%

\begin{figure*}
\epsscale{1}
\plotone{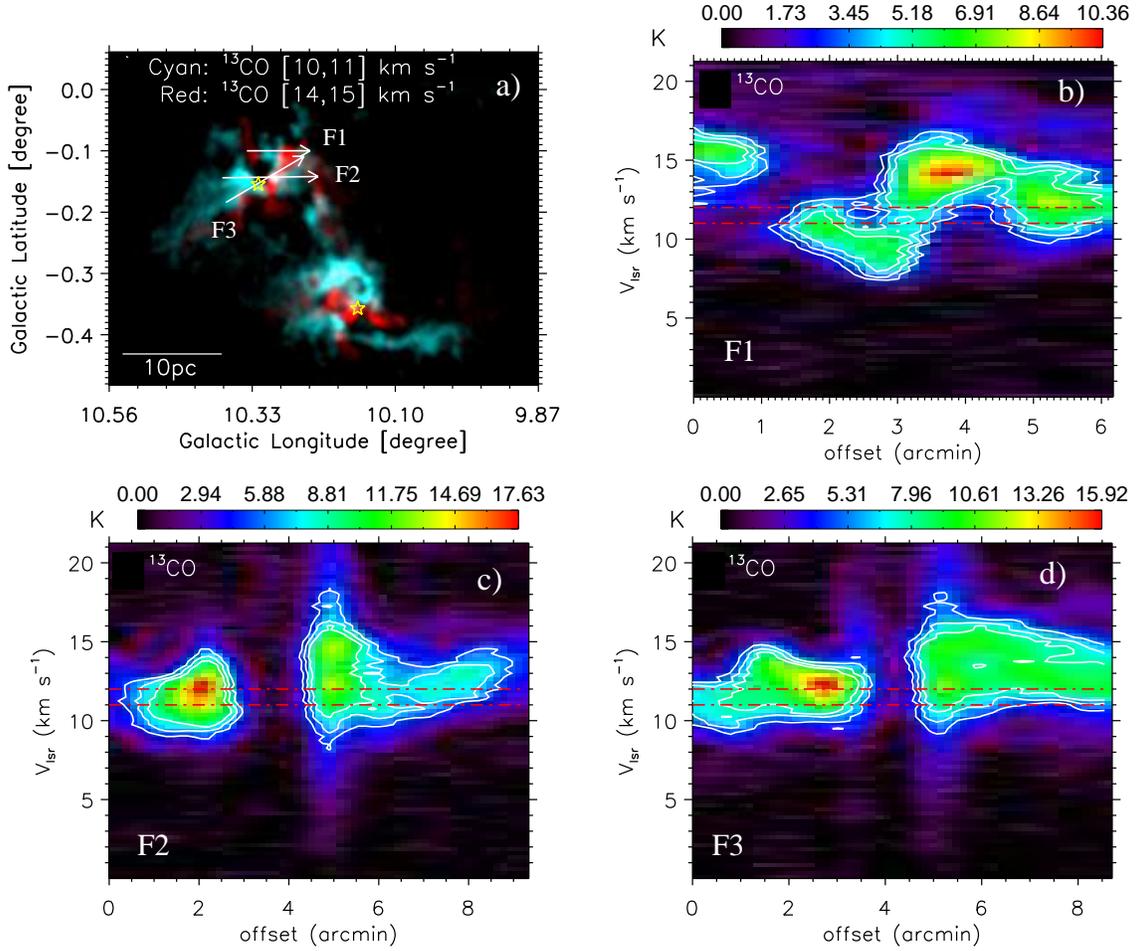}
\caption{a) The panel shows a two-color composite image made using the SEDIGISM {\tco} map at [14, 15] {\kps} (in red) and the SEDIGISM {\tco} map at [10, 11] {\kps} (in cyan) toward the W31 complex. 
Stars are the same as marked in Figure~\ref{fig1}a. Three arrows ``F1", ``F2", and ``F3" are marked in the direction of W31-N, where the position-velocity diagrams are produced (see Figures~\ref{fig11}b--\ref{fig11}d). Position-velocity diagrams along the arrows b) ``F1''; c) ``F2''; and d) ``F3'', respectively as shown in Figure~\ref{fig11}a. 
Two dotted dashed lines (in red) at V$_\mathrm{lsr}$ = 11 and 12 {\kps} are marked in each position-velocity diagram.}
\label{fig11}
\end{figure*}

Based on the SEDIGISM {\tco} moment-1 map, a noticeable velocity gradient is also seen toward W31-N (see Figures~\ref{fig3}e and~\ref{fig3}f), hinting at the presence of similar cloud components toward W31-N as observed in the direction of W31-S.  In Figure~\ref{fig11}a, we present the same two color-composite image (using the {\tco} maps at [10, 11] and [14, 15] {\kps}) as shown in Figure~\ref{fig9}a, but covering both the sites W31-N and W31-S. The position-velocity diagrams of {\twco} line data toward W31-N do not show a clear signature of two cloud components (see Section \ref{subsec:sfc}).  Using the {\tco} data, Figures~\ref{fig11}b,~\ref{fig11}c, and~\ref{fig11}d present the position-velocity diagrams along the arrows ``F1'', ``F2'' and ``F3'', respectively (as indicated in Figure~\ref{fig11}a). In Figure~\ref{fig11}b, there is a hint of the presence of two velocity components, but it is not convincing as seen toward W31-S.  
It is quite possible that the signature of CCC in W31-N is diminished by the feedback of O-type stars, which is evident by the presence of MIR bubble CN 148.
Taking into account the age difference between young protostars and {\htwo} regions, there was a suggestion of triggered star formation through the feedback of massive O-type stars in W31-N \citep[e.g.,][]{Beuther, CN148}. 
Therefore, based on our observations, we can say that the possibility of CCC cannot be completely ruled out in W31-N.
This paper does not focus on the impact of massive stars to their surroundings. Hence such study is beyond the scope of this paper.

Furthermore, the outcomes of smoothed particle hydrodynamics simulations concerning the head-on collision of two clouds result in the formation of filaments \citep{balfour15}.  According to this theoretical work, the colliding clouds generate a shock-compressed layer, which fragments into the filaments.  With lower velocities of collision, the lateral collapse of the layer drags and stretches the filaments towards the collision axis, making the filaments predominantly radial.  However, in the case of higher collision velocities, less time for the lateral collapse leads to the formation of filaments like a spider's web.  A recent review article on CCC by \citet{fukui21} also agrees well with \citet{balfour15} to demonstrate the role of collision in forming filamentary structures.  Observationally, \citet{beltran22} studied a massive star-forming region and found a deep connection between CCC in the formation of HFS and massive protocluster associated with the hot molecular core at G31.41+0.31 using the N$_{2}$H$^{+}$(1--0) observations.  In addition, the work of \citet{dewangan22ex} on the infrared dark cloud (IRDC) G333.73+0.37 based on the SEDIGISM $^{13}$CO and C$^{18}$O line data and the study of \citet{fukui19ex} on N159E-Papillon Nebula located in the Large Magellanic Cloud (distance $\sim$50 kpc) provided observational results in favor of CCC or converging flows, explaining the presence of the HFS and massive stars.  The observational findings of our study on the W31 complex also greatly match the prescriptions derived from the works connecting CCC and the formation of HFSs. Hence, our proposed collision process may also explain the existence of HFSs and the formation of massive stars in our target area. The presence of HFSs suggests the gas inflow from the large scale to the central hubs. In this way, the clump-fed scenario (see Section~\ref{sec:intro}) of massive star formation is also applicable in W31. To infer the gas motion along these filaments (see Figure~\ref{fig2}), one requires the information of dense gas kinematics, which is one of the observational limitations of our current work.

\section{Summary and Conclusions}
\label{sec:conc}
The present paper focuses on unveiling the physical process concerning the formation of O-type stars observationally in the W31 complex. ln this relation, we employ multi-wavelength data sets in the direction of the W31 complex, which include new NANTEN2 {\twco} and archival SEDIGISM {\tco} data. The main findings of the present study are summarized below.
\begin{enumerate}
\item A total of 49 ATLASGAL 870 $\mu$m dust continuum clumps and several Class~I protostars are spatially traced toward W31-N and W31-S. Out of 49, 25 ATLASGAL clumps are associated with molecular outflow activity hence, support the ongoing star formation processes. These dust clumps in a velocity range of [8.9, 15.3] {\kps} are situated at d = 3.55 kpc, illustrating that W31-N and W31-S belong to a single extended physical system.
\item Based on the {\it Herschel} 250 {\micro} image, {\it{getsf}} reveals several filaments toward W31-N and W31-S. In both W31-N and W31-S, filaments are found to converge at the central area of the {\htwo} regions.  These findings support the presence of HFSs in both the regions W31-N and W31-S.

\item The observed complementary distribution suggests a small inclination of the viewing axis to the axis of collision, which is considered to be about 10{\degr}.  The size of the overlapping zone of the cloud components along the collision axis and a relative velocity is computed to be {\simi}14.3 pc and {\simi}8.1 {\kps}, respectively.  These parameters yield a collision time-scale of {\simi}2 Myr. 
%
\item Signposts of star formation, including massive O-type stars in W31-S, are concentrated toward the overlapping areas of the two cloud components.  The derived collision time-scale is sufficiently old enough to influence massive star formation in the W31 complex.
\item In the direction of W31-N, the signatures of CCC are not as promising as W31-S.
\end{enumerate} 


Overall, our observational results support the CCC scenario in W31 complex, which triggers the formation of HFSs that host O-type stars at the central hubs.
The filaments further accrete material from the large scale to feed the dense hub, favoring the clump-fed scenario of massive star formation.
\section*{Acknowledgments}
We thank the referee for valuable comments and suggestions.
The research work at Physical Research Laboratory is funded by the Department of Space, Government of India. 
NANTEN2 is an international collaboration of 11 universities: Nagoya University, Osaka Prefecture University, University of Bonn, University of Cologne, Seoul National University, University of Chile, University of New South Wales, Macquarie University, University of Sydney, University of Adelaide, and the University of ETH Zurich.
This publication is based on data acquired with the Atacama Pathfinder Experiment (APEX) under programs 092.F-9315 and 193.C-0584.  APEX collaborates with the Max-Planck-Institut fur Radioastronomie, the European Southern Observatory, and the Onsala Space Observatory.  The processed data products are available from the SEDIGISM survey database located at https://sedigism.mpifr-bonn.mpg.de/index.html, constructed by James Urquhart and hosted by the Max Planck Institute for Radio Astronomy. This research made use of Astropy\footnote{http://www.astropy.org}, a community-developed core Python package for Astronomy \citep{astropy13,astropy18}.
\par


\bibliographystyle{aasjournal}
\bibliography{reference}{}

%
%

%
%

\end{document}